THE AHARONOV-BOHM EFFECT IN THE MOMENTUM SPACE

D. Dragoman*, S. Bogdan – Univ. Bucharest, Physics Dept., P.O. Box MG-11, 077125

Bucuresti-Magurele, Romania

ABSTRACT:

The Schrödinger formalism of quantum mechanics is used to demonstrate the existence of the

Aharonov-Bohm effect in momentum space and set-ups for experimentally demonstrating it are proposed for either free or ballistic electrons.

INTRODUCTION

The Aharonov-Bohm effect [1] illustrates the influence of the vector potential of the magnetic field upon the motion of charged particles, which alters the phase of the wavefunction of charged carriers although no net Lorentz force acts upon these carriers. In particular, the Aharonov-Bohm effect produces a shift of the interference fringes inside the unchanged region of interference of two coherent beams of charged particles. Countless experimental observations of the Aharonov-Bohm effect were performed since its discovery, most of them involving electrons; typical Aharonov-Bohm and related experiments are examined in Ref.2. However, the result of all these experiments can be interpreted by treating the electron wavefunction in the configuration space.

Only recently it has been demonstrated that a similar Aharonov-Bohm effect exists also in momentum space [3]: the interference fringes of overlapping electron wavefunctions in the momentum space are shifted by the vector potential $A$ even if the magnetic field $B = \nabla \times A$ vanishes across the spatial region traversed by the interfering electron beams. The theoretical demonstration of the existence of the Aharonov-Bohm effect in the momentum space was done employing the phase space formalism of quantum mechanics; according to the authors' knowledge no attempt to experimentally evidence this effect was made.

After briefly demonstrating the existence of the Aharonov-Bohm effect in momentum space using the Schrödinger formulation of quantum mechanics, this paper proposes several setups that are able to evidence this effect experimentally. It is important to note that the existence of the Aharonov-Bohm effect in the momentum space does not follow immediately from its existence in the configuration space. As will be shown below (see also Ref.4), not even the interference in momentum space follows directly from the existence of interference in the configuration space; it is possible to have interference in either momentum or configuration space, in both spaces or in none.

THE AHARONOV-BOHM EFFECT IN MOMENTUM SPACE

For the simplicity of the analytical treatment let us consider the situation represented in Fig.1: a coherent electron beam passes through two slits, the two outgoing electron beams propagating in a region in which the magnetic field vanishes ($\boldsymbol{B} = 0$) and the vector potential is different from zero ($\boldsymbol{A} \neq 0$) in such a way that the two electron beams encompass a region in which the total flux $\Phi \neq 0$. This condition can be realized if the current-carrying solenoid placed immediately after the slits and centered with respect to them is tightly wound.

The total electron wavefunction in the configuration space, $\Psi(\boldsymbol{r}, t) = \Psi_1(\boldsymbol{r}, t) + \Psi_2(\boldsymbol{r}, t)$, has two components corresponding to the two interfering electron beams and evolves according to the Hamiltonian $H = (\boldsymbol{p} - e\boldsymbol{A})^2 / 2m$, with $\boldsymbol{p}$ the electron momentum, $m$ its mass and $e$ the electrical charge. The general solution for the electron wavefunction takes the form $\Psi_1 = \Psi_1^0 \exp(-i\boldsymbol{j}_1)$, $\Psi_2 = \Psi_2^0 \exp(-i\boldsymbol{j}_2)$, where the phase difference is given by $\Delta\boldsymbol{j} = \boldsymbol{j}_1 - \boldsymbol{j}_2 = (e/\hbar)\oint \boldsymbol{A}d\boldsymbol{r} = 2\boldsymbol{p}(\Phi/\Phi_0)$, $\Phi = \oint \boldsymbol{A}d\boldsymbol{r}$ is the magnetic flux, in which the integral is performed along the boundaries of the region delimited by the electron trajectories, and $\Phi_0 = h/e$ is the quantum of the magnetic flux. Equivalently, if the electron beams interact with the solenoid during a small time interval, the situation represented in Fig.1 can be modeled as the propagation of two coherent electron beams, which evolve most of the time freely, i.e. with $\boldsymbol{A} = 0$ under a Hamiltonian $H = \boldsymbol{p}^2 / 2m$, the influence of the solenoid manifesting itself in the inclusion of phase factors $\pm \Delta\boldsymbol{j} / 2$ in the wavefunctions of the interfering beams immediately after the plane of the slits.

More precisely, let us consider for simplicity a one-dimensional normalized total wavefunction, which at $t = 0$, i.e. immediately after the slits, is given in the configuration space by

$$\Psi(x,0) = \Psi_1(x,0) + \Psi_2(x,0) = N[e^{-(x+d)^2/x_0^2}e^{-i\Delta\boldsymbol{j}\,/2} + e^{-(x-d)^2/x_0^2}e^{i\Delta\boldsymbol{j}\,/2}].$$ (1)

Here we assume Gaussian slits with extension $x_0$ and separated by a distance of $2d$, the normalization factor being $N = [\sqrt{2\boldsymbol{p}}x_0(1 + e^{-2d^2/x_0^2}\cos\Delta\boldsymbol{j}\,)]^{-1/2}$. This initial wavefunction in the configuration space consists of two non-overlapping terms, which represent the electron wavefunctions immediately after the slits; $|\Psi(x,0)|^2$ is represented in Fig.2a for $\Delta\boldsymbol{j}\,=\boldsymbol{p}\,/2$ and $d = 4x_0$.

The initial wavefunction in the momentum space can be calculated from

$$\Theta(p,0) = (2\boldsymbol{p}\hbar)^{-1/2}\int_{-\infty}^{+\infty}\exp(-ipx/\hbar)\Psi(x,0)dx$$ (2)

and is given by

$$\Theta(p,0) = N'e^{-\frac{p^2 x_0^2}{4\hbar^2}}\cos\left(\frac{\Delta\boldsymbol{j}}{2} - \frac{dp}{\hbar}\right),$$ (3)

where $N' = Nx_0(2/\hbar)^{1/2}$. As can be seen from the probability distribution in the momentum space, $|\Theta(p,0)|^2$, which is represented in Fig.2b for $\Delta\boldsymbol{j}\,=\boldsymbol{p}\,/2$, interference fringes appear in $p$ at $t = 0$, although the interfering electron beams do not spatially overlap. So, interference in the momentum space does not necessarily imply interference in the configuration space and vice-versa; this fact was already pointed out in Ref.4. Analogously, the existence of the Aharonov-Bohm effect in configuration space does not necessarily imply a similar effect in momentum space. But, it can be observed that the interference fringes in $p$ are shifted from the $A = 0$ case, when the probability distributions in the configuration and momentum spaces look like in Figs.2c and 2d, respectively. These probability distributions are given by the same

expressions as in (1) and (3), respectively, with $\Delta \boldsymbol{j} = 0$. The shift of the interference fringes in the momentum space when $\Phi \neq 0$ (when $A \neq 0$, $\Delta \boldsymbol{j} = 0$) but $B = 0$ is the signature of the Aharonov-Bohm effect in momentum space.

The interference fringes in momentum space remain unchanged at free propagation after the plane of the slits, under the action of the Hamiltonian $H = p^2 / 2m$, since the wavefunction in the momentum space evolves in time according to

$\Theta(p,t) = \Theta(p,0) \exp(-ip^2 t / 2m\hbar)$, and hence $|\Theta(p,t)|^2 = |\Theta(p,0)|^2$.

The free space evolution only modifies the wavefunction in configuration space, which for $t > 0$ is given by

$$\Psi(x,t) = N'' e^{-\frac{(x^2+d^2)(x_0^2-2it\hbar/m)}{x_0^4+(2t\hbar/m)^2}} \cos\left[\frac{\Delta \boldsymbol{j}}{2} - \frac{2xd(ix_0^2 + 2t\hbar/m)}{x_0^4 + (2t\hbar/m)^2}\right], \qquad (4)$$

with $N'' = N' (2\boldsymbol{p}\hbar)^{-1/2} [4\boldsymbol{p}m\hbar^2 /(x_0^2 m + 2it\hbar)]^{1/2}$. This expression is obtained using the inverse Fourier transform

$$\Psi(x,t) = (2\boldsymbol{p}\hbar)^{-1/2} \int_{-\infty}^{+\infty} \exp(ipx/\hbar)\Theta(p,t)dp . \qquad (5)$$

The probability densities in the configuration and momentum spaces after a free propagation time $t = 5mx_0^2 /(2\hbar)$ are represented in Figs.3a and 3b when $\Delta \boldsymbol{j} = \boldsymbol{p} / 2$, whereas in Figs.3c and 3d the same graphics are drawn for $\Delta \boldsymbol{j} = 0$. The difference from Fig.2 is that interference appears now also in the configuration space since the interfering coherent electron beams begin to overlap spatially. The Aharonov-Bohm effect is now present also in configuration space, the interference fringes in Fig.3a being shifted from the position in the absence of the vector potential (see Fig.3c), when $\Delta \boldsymbol{j} = 0$.

PROPOSED SET-UPS FOR EXPERIMENTAL OBSERVATION OF THE AHARONOV-BOHM EFFECT IN MOMENTUM SPACE

Can the interference in momentum space be evidenced experimentally? The answer is affirmative: interference in momentum space for neutrons has been observed with the help of a skew-symmetric perfect crystal neutron interferometer [4] followed by an analyzer crystal that measures the spectral distribution of one of the two outgoing neutron beams, which consists in its turn of two spatially non-overlapping components. This set-up cannot be directly used to observe the Aharonov-Bohm effect in momentum space since neutrons have no electric charge and hence the vector potential does not influence their movement. However, a similar interferometric set-up for electrons can be used instead.

This set-up, which can evidence the Aharonov-Bohm effect in momentum space for free electrons, is represented in Fig.4. It is a Mach-Zender-type interferometer consisting of beam splitters for electrons $D_1$ and $D_2$ implemented through thin crystalline lamellae [5] and mirrors for electrons, $M_1$ and $M_2$, which can be for example intense laser beams [6]. This interferometer must be supplemented by a spectrometer, as in the case of the set-up in [4] for neutrons, to observe the Aharonov-Bohm effect in $p$ when the solenoid S carries electrical current.

Another set-up for free electrons, which does not rely on a spectrometer for electrons, is represented in Fig.5a. This set-up is similar to that employed in Ref.7, but followed by an electron lens with focal length $f$, which implements at a distance $f$ after the lens the Fourier transform of a wavefunction incident at a distance $f$ before the lens [8]. An electron lens [9] is for example a magnetic quadrupole; such a lens that focalizes a two-dimensional electron distribution is represented in Fig.5b. The confined magnetic field with $A \neq 0$ is generated by a toroidal permalloy probe covered by a superconductor Nb layer. The electron biprism consists of two planar electrodes and a filament [7].

It should be also possible to observe the Aharonov-Bohm effect in momentum space for ballistic electrons, in a two-dimensional electron gas (2DEG). A possible set-up, represented in Fig.6, makes use of a beam of ballistic electrons that propagates in an electron waveguide and is then split in two beams, which are subsequently reunited in an interference region F after propagating along different paths that encompass the solenoid S. The ballistic electron wavefunction in configuration space, in the interference region F, can be subsequently Fourier transformed in order to obtain the wavefunction in momentum space by applying a magnetic field $B$ parallel to the 2DEG plane [10]. This nonvanishing magnetic field should not be confused with the vanishing magnetic field, which produces a nonvanishing vector potential, that generates the Aharonov-Bohm effect; it is applied in the region that immediately follows the formation of interference fringes. The Fourier transform is obtained after a specific propagation length $L$, which depends on the applied parallel magnetic field. It is worth noting that a set-up similar to that in Fig.6, except the Fourier-transforming part, was used in Ref.11 to demonstrate the Aharonov-Bohm effect in the configuration space. Alternativly, the Fourier-transforming part can be replaced by a simple electron lens with focal length $f$, which (as mentioned above) implements at a distance $f$ after the lens the Fourier transform of a wavefunction incident at a distance $f$ before the lens. Both electrostatic and magnetic lenses have been demonstrated for ballistic electrons (see [12] and the references therein).

CONCLUSIONS

In conclusion, we have shown the existence of the Aharonov-Bohm effect in momentum space using the Schrödinger formalism of quantum mechanics, and have proposed set-ups able to demonstrate experimentally this effect for either free or ballistic electrons. To the best of authors' knowledge the Aharonov-Bohm effect in momentum space has not been experimentally observed up to now, although all the set-ups that we have proposed can be realized using present technology.

The conclusions of the present paper, in particular the existence of a similar effect in momentum space, can be extended to effects related to the Aharonov-Bohm such as [13] the electric Aharonov-Bohm effect, the Aharonov-Casher effect, or the scalar Aharonov-Bohm effect.

FIGURE CAPTIONS

Fig.1    Geometry of the experimental set-up that can be used to evidence the Aharonov-Bohm effect in configuration space

Fig.2    Probability densities in (a) configuration space and (b) momentum space immediately the slits for $d = 4x_0$, $\Delta \boldsymbol{j} = \boldsymbol{p} / 2$. (c), (d) Idem as in (a), (b) for $\Delta \boldsymbol{j} = 0$

Fig.3    Same as in Fig.2 after a free propagation time $t = 5mx_0^2 / (2\hbar)$

Fig.4    Experimental set-up able to demonstrate the Aharonov-Bohm effect in momentum space for free electrons

Fig.5 (a) Another experimental set-up able to demonstrate the Aharonov-Bohm in momentum space for free electrons. (b) An electron lens.

Fig.6    Experimental set-up able to evidence the Aharonov-Bohm effect in momentum space for balistic electrons

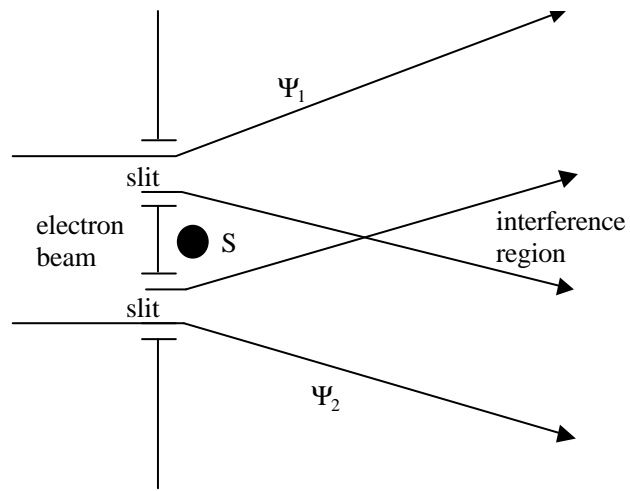

Fig.1

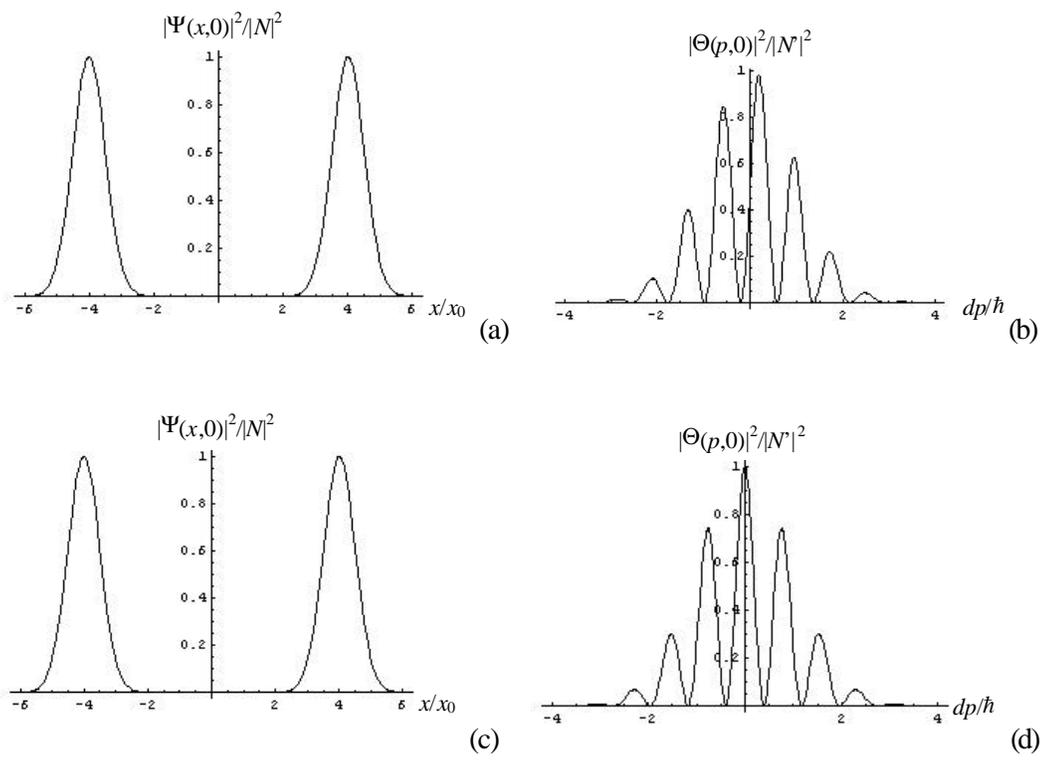

Fig.2

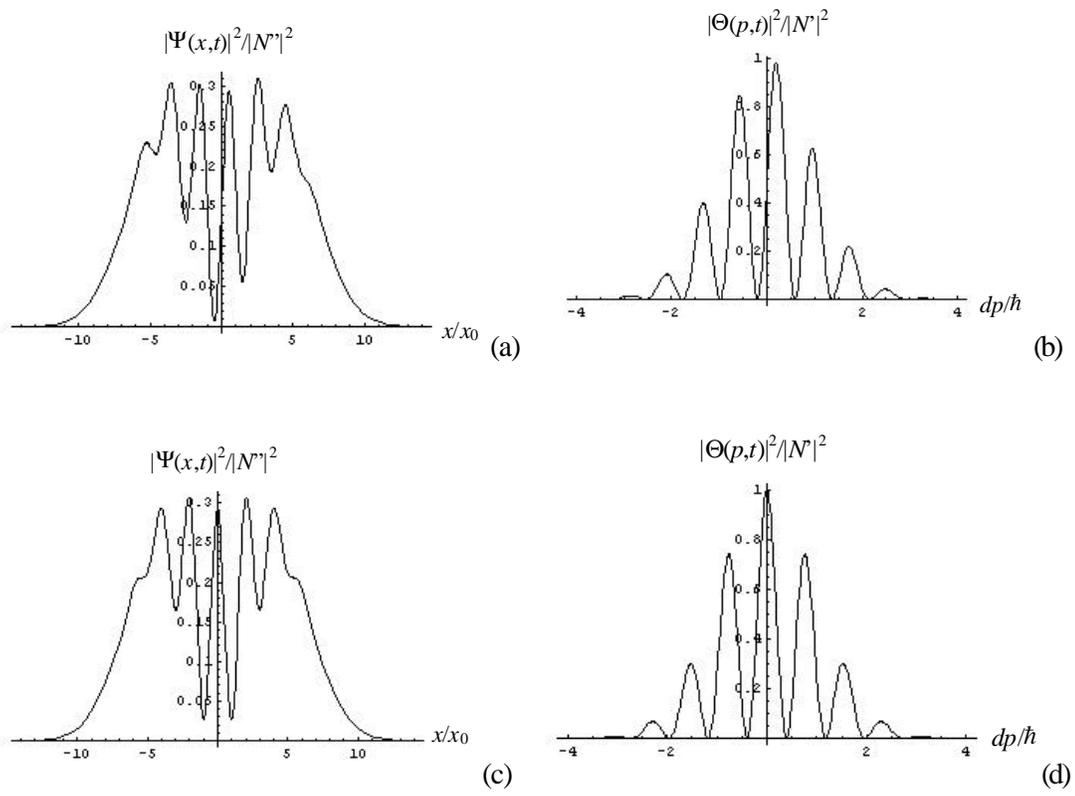

$|\Psi(x,t)|^2/|N'|^2$

$|\Theta(p,t)|^2/|N'|^2$

$x/x_0$ (a)

$dp/\hbar$ (b)

$|\Psi(x,t)|^2/|N'|^2$

$|\Theta(p,t)|^2/|N'|^2$

$x/x_0$ (c)

$dp/\hbar$ (d)

Fig.3

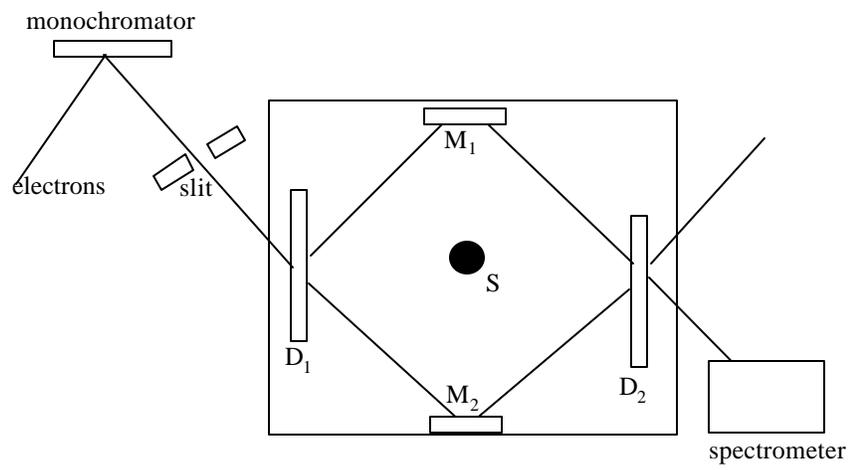

monochromator

electrons

slit

M₁

S

D₁

M₂

D₂

spectrometer

Fig.4

reference
wave

probe

object
wave

lens

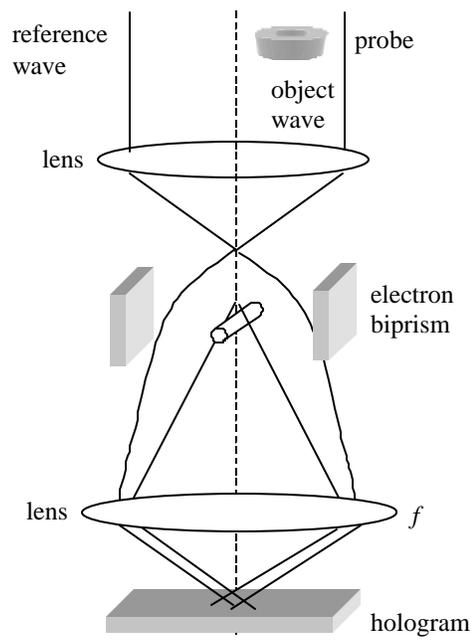

electron
biprism

lens   $f$

hologram

(a)

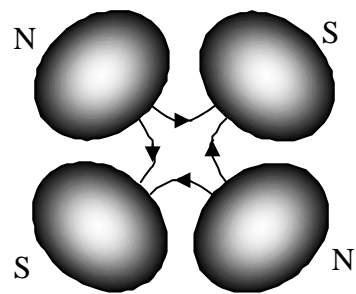

N        S

S        N

(b)

Fig.5

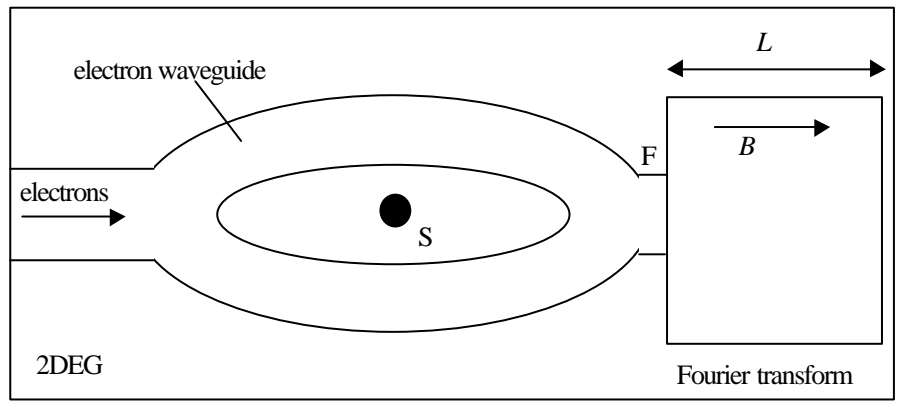